\documentclass[12pt,letterpaper]{article}

\usepackage{hyperref}

\usepackage{amsmath}
\usepackage{amsthm}
\usepackage{amssymb}
\usepackage{graphicx}

\newtheorem{thm}{Theorem}

\newtheorem{defn}{Definition}
\newtheorem{obs}{Observation}

\author{
Hyung-Chan An
\thanks{
{\tt hyung-chan.an@yonsei.ac.kr}.
Corresponding author.
Department of Computer Science, Yonsei University, Seoul 03722, South Korea.
Research supported in part by NSF under grants no. CCF-1017688 and CCF-0729102, and the Korea Foundation for Advanced Studies.
Part of this research was conducted while the author was a PhD student at Cornell University.
This work was supported by the National Research Foundation of Korea (NRF) grant funded by the Korea government (MSIT) (No. NRF-2019R1C1C1008934).
}
\and
Robert Kleinberg
\thanks{
{\tt rdk@cs.cornell.edu}.
Department of Computer Science, Cornell University, Ithaca, NY 14853.
Supported by NSF grants CCF-0643934 and CCF-0729102, AFOSR grant FA9550-09-1-0100, a Microsoft Research New
Faculty Fellowship, a Google Research Grant, and an Alfred P. Sloan Foundation Fellowship.
}
}
\title{A Diameter-Revealing Proof of the Bondy-Lov\'{a}sz Lemma\footnote{Link to the formal publication: \url{https://doi.org/10.1016/j.ipl.2021.106194}}}
\date{}

\begin{document}

\maketitle

\begin{abstract}
We present a strengthened version of a lemma due to Bondy and Lov\'{a}sz. This lemma establishes the connectivity of a certain graph whose nodes correspond to the spanning trees of a 2-vertex-connected graph, and implies the $k=2$ case of the Gy\H{o}ri-Lov\'{a}sz Theorem on partitioning of $k$-vertex-connected graphs. Our strengthened version constructively proves an asymptotically tight $O(|V|^2)$ bound on the worst-case diameter of this graph of spanning trees.
\end{abstract}

\section{Introduction}

The Gy\H{o}ri-Lov\'{a}sz Theorem~\cite{G,L} asserts that a $k$-vertex-connected graph $G=(V,E)$, for any distinct $u_1,\ldots,u_k\in V$ and $n_1+\cdots+n_k=|V|$, can be partitioned into $k$ vertex-disjoint connected subgraphs where the $i$-th subgraph consists of exactly $n_i$ vertices including $u_i$.
In the case $k=2$, Lov\'{a}sz~\cite{L} provided an elegant proof based on a lemma due to Bondy and Lov\'{a}sz that a certain graph (of exponential or even superexponential size)
is connected.  The vertices of this graph are the spanning
trees of $G$; for a specified vertex $a\in V$, two spanning trees
are adjacent if their intersection contains a tree on
$\mbox{$|V|-1$}$ vertices including $a$.
The proof in~\cite{L} establishes only an exponential upper bound
on the diameter of this graph, leaving unresolved the question of
whether the graph has polynomial diameter.

In this paper, we present a strengthened version of the Bondy-Lov\'{a}sz lemma that constructively proves an $O(|V|^2)$ bound on the worst-case diameter of this graph of spanning trees.
We also show that this bound is asymptotically tight.

\paragraph{Algorithmic motivation for our results.}
One motivation for our results stems from the challenge of understanding
the computational complexity of the
\emph{Gy\H{o}ri-Lov\'{a}sz Search Problem}: given a $k$-vertex-connected
graph, find a spanning forest composed of $k$ trees with specified root vertices
and sizes. This problem is known to be solvable in polynomial time
when $k=2$~\cite{L} or $k=3$~\cite{STNMU}, but for $k > 3$ it is only known
to belong to the complexity class PLS~\cite{chandran2018spanning}.
Lov\'{a}sz's polynomial-time algorithm in the case $k=2$ stems from his
proof that the graph of spanning trees is connected: the method of proof
yields a polynomial-time algorithm that essentially performs bisection
search\footnote{The algorithm iteratively splits the path into two
subpaths and recurses on one of the subpaths, but unlike in bisection search,
the two subpaths are not necessarily of equal size. Nevertheless
a different progress
measure can be used to prove that the number of iterations of the search
process is at most the number of vertices of $G$.}
on the (potentially exponentially long) path linking two spanning
trees. Our quadratic upper bound on the diameter of the graph of
spanning trees yields a different algorithm
for the $k=2$ case of the Gy\H{o}ri-Lov\'{a}sz Search Problem,
based on a sequential search of a polynomially long path.
The algorithm defines $G^+$ to be the 2-vertex-connected
graph obtained from the given graph $G$ by adding a vertex $a$
and edges $(u_1,a)$ and $(u_2,a)$. For $i \in \{1,2\}$ let
$T_i$ be a spanning tree of $G^+$ obtained by deleting the
vertex $u_i$ from $G^+$, taking any spanning tree of the
resulting graph, and reattaching $u_i$ as a leaf of that tree.
By Theorem~\ref{thm:ub} below, there is a polynomial-time
algorithm to compute a path $P$ in the graph of spanning trees
of $G^+$ (rooted at $a$) such that $P$
starts at $T_1$, ends at $T_2$, and has length
$O(|V|^2)$.
For any spanning tree of $G^+$ rooted at $a$, let $N_1(T)$ denote
the number of vertices in the subtree rooted at $u_1$, excluding
$u_2$ and its descendants. Any pair of adjacent
trees $T,T'$ satisfy $|N_1(T) - N_1(T')| \le 1$. Since
$N_1(T_1) = 1$ and $N_1(T_2) = |V(G)|$, as $T$ ranges over the
trees in path $P$ the value $N_1(T)$ must take every value
in the range $\{1,2,\ldots,n\}$. Therefore, a brute-force
search of the $O(|V|^2)$ trees that constitute $P$ is assured of finding
a tree $T$ with $N_1(T) = n_1.$ Deleting $a$ from $T$,
and disconnecting $u_2$ from its parent if that parent is not $a$,
one obtains a spanning forest of $G$ whose two components have
sizes $n_1$ and $|V(G)|-n_1$ and roots $u_1$ and $u_2$, respectively.

For $k>2$, Lov\'{a}sz's topological proof~\cite{L} of the
Gy\H{o}ri-Lov\'{a}sz Theorem is based on constructing
a topological space 
that generalizes the graph of spanning trees used in the
$k=2$ case and satisfies a topological connectivity
property, defined in terms of reduced
homology groups, that generalizes the connectedness of the
graph of spanning trees. (See \cite[Theorem 29]{10.1145/1255443.1255444}
for a precise formulation of the relevant topological connectivity
property.) Lov\'{a}sz's proof does not lead directly to a
polynomial-time algorithm because the topological space
defined in the proof is composed of a potentially (super)exponential
number of polyhedral cells.
Unlike in the $k=2$ case, it is not known whether bisection search
(or a higher-dimensional generalization thereof) can be
used to search this \mbox{(super)exponentially} large cell complex in
polynomial time. However, if the cell complex could be
``sparsified'' in polynomial time by extracting
a subcomplex, composed of only polynomially many cells,
that satisfies the same topological connectivity
property as in Lov\'{a}sz's proof, then brute-force
search over the vertices of that subcomplex would solve
the Gy\H{o}ri-Lov\'{a}sz Search Problem in polynomial time.
Our Theorem~\ref{thm:ub} implements this computationally
efficient sparsification procedure when $k=2$; the
subcomplex in that case is the path $P$ defined above.
We hope this may motivate investigation into the existence
of efficient sparsification procedures when $k>2$,
although constructing such a sparsification,
if it is even possible, would almost assuredly
require more sophisticated mathematics than the methods
deployed in the proof of Theorem~\ref{thm:ub}.

\paragraph{Related work.}
There  exist alternative proofs of (generalizations of) the Gy\H{o}ri-Lov\'{a}sz Theorem.
Hoyer and Thomas~\cite{hoyer2016gyorilovasz} presented an alternative exposition of Gy\H{o}ri's proof; Idzik~\cite{IDZIK1990297} presented a proof in the same spirit as Gy\H{o}ri's to give a slightly stronger conclusion: given a partition of $G$ into connected subgraphs $V_1,\ldots,V_k$ each containing $u_1,\ldots,u_k$, if $V_1$ has more than one vertex, then there is another partition $V'_1,\ldots,V'_k$ (again, each containing $u_1,\ldots,u_k$) such that $V'_1$ has one fewer vertices than $V_1$, $V'_k$ is a proper superset of $V_k$, and $|V'_i|=|V_i|$ for all $i=2,\ldots,k-1$. Chen et al.~\cite{10.1145/1255443.1255444} proved a version with vertex weights by generalizing Lov\'{a}sz's topological proof; Chandran et al.~\cite{chandran2018spanning}, among other results, rederived the vertex-weighted generalization using a proof similar to Gy\H{o}ri's, obtaining an $O^*(4^n)$-time algorithm for constructing the partition.

\section{Upper bound}

\begin{defn}
For $G=(V,E)$ with a specified vertex $a\in V$, two spanning trees of $G$ are \emph{adjacent} if their intersection contains a tree on $|V|-1$ vertices including $a$.
\end{defn}

From now on, we will consider spanning trees as rooted at $a$. Let $n:=|V|$. We assume that $n\geq 2$.

\begin{obs}
Two spanning trees $T_A$, $T_B$ are adjacent if and only if $T_B$ can be obtained by detaching some leaf $v\neq a$ of $T_A$ from its current parent and attaching it to some vertex.
\end{obs}

\begin{thm}\label{thm:ub}
Let $G=(V,E)$ be a 2-vertex-connected graph and let $a$ be a specified vertex
of $G$. For any two spanning trees $T,T'$ of $G$, there is a sequence of
at most $O(n^2)$ trees beginning with $T$ and ending with $T'$,
such that every pair of consecutive trees in the
sequence are adjacent.
Moreover, this path can be found in polynomial time.
\end{thm}
\begin{proof}

Recall that an $st$-numbering of a graph $G$ with respect to
an edge $(s,t)$ is a numbering of the vertices of $G$ as
$v_1,\ldots,v_n$ such that $s=v_1, t=v_n$, and every vertex
$v_i \neq s,t$ has two neighbors $v_j,v_k$ such that $j < i < k$.
It is well-known that every 2-vertex-connected graph
has an $st$-numbering with respect to every one of its
edges~\cite{LEC}.  Let us choose an arbitrary edge incident to the
distinguished vertex $a$, and
let  $v_1,v_2,\ldots,v_n$ be an $st$-numbering with
respect to this edge, such that $v_1=a$.
The $st$-numbering can be found in polynomial time~\cite{LEC,even1976computing,ebert1983st,tarjan1986two,brandes2002eager}.

Let $T^+$ be the ``canonical'' spanning tree constructed as follows: $v_n$ is a child of $v_1$; every vertex other than $v_1$ and $v_n$ is a child of its highest-numbered neighbor. It is easy to show by induction that $T^+$ is a uniquely defined spanning tree.

It suffices to prove the theorem only for $T=T^+$. In constructing a sequence of spanning trees beginning with $T^+$ and ending with an arbitrary spanning tree $T'$, we identify ``milestones'' $T_1=T^+,T_2,\ldots,T_{n-1},T_n=T'$ where each pair of consecutive milestones are joined by a sequence of  $O(n)$ spanning trees, each adjacent to the next one in the sequence. First, we define $S_1,\ldots, S_n$ that are connected subgraphs of $T'$ containing $a$. Note that the vertex set of $S_k$, $V(S_k)$, uniquely determines $S_k$. 
Our construction will satisfy $S_1\subsetneq\cdots\subsetneq S_n$, where $S_1$ is the singleton tree $\{a\}$, $S_{k+1}$ contains $S_k$ and one other vertex, and $S_n=T'$. In particular, among all $(u,v)\in T'$ such that $u\in S_k$ and $v\notin S_k $, choose $(u^*_k,v^*_k)$ in which $v^*_k$ has the highest number; $V(S_{k+1}):=V(S_k)\cup\{v^*_k\}$.

The spanning tree $T_k$ is defined to be a supergraph of $S_k$. In $T_k$, every vertex $v$ in $V\setminus V(S_k)$ becomes a child of its highest-numbered neighbor unless $v=v_n$. If $v=v_n$, $v$ becomes a child of $v_1$. It is easy to see that $T_k$ is indeed a tree. We have $T_1=T^+$ and $T_n=S_n=T'$.

Now, for $1\leq k<n$, we present an algorithm that produces a sequence of $O(n)$ spanning trees beginning with $T_k$ and ending with $T_{k+1}$ such that every pair of consecutive trees are adjacent. First, for each $v\in V\setminus V(S_k)$ in the ascending order of the $st$-numbering, if $v\neq v_k^*$, we detach $v$ from its current parent and attach it to its lowest-numbered neighbor; if $v=v^*_k$, we detach $v$ from its current parent, attach it to $u^*_k$, and stop processing further vertices in $V\setminus V(S_k)$. Note that $v\neq v_n$ in the first case. Then, for every vertex $v$ that was reattached in the first loop except for the last one $v^*_k$, in the reverse order (i.e., descending order of the numbering), detach $v$ from its current parent and attach it to its highest-numbered neighbor. The algorithm outputs the snapshot of the current spanning tree after each reattachment.

We claim that every vertex that was reattached during this process was a leaf at the time of detachment; then, this algorithm produces a sequence of $O(n)$ spanning trees where every pair of consecutive spanning trees are adjacent. All of these $O(n)$ spanning trees contain $S_k$, because the vertices in $S_k$ are never detached by the algorithm. In the second loop, every detached vertex is attached back to its parent in $T^+$, except for $v^*_k$ that is now attached to $u_k^*$; thus, the last spanning tree produced by the algorithm is $T_{k+1}$.

To complete the proof, it remains to verify the claim that every vertex $v$ that was reattached during this process was a leaf at the time of detachment. We implicitly use induction on the number of iterations of the algorithm.

In an iteration of the first loop, suppose $v_i$ gets reattached but was not a leaf. Let $v_j$ be its arbitrary child in the tree before the reattachment. Observe that $v_j\notin S_k$, since $v_i\notin S_k$ and $S_k$ is a connected subtree contained in all the spanning trees. Suppose $j>i$; then $v_j$ has not been considered by the algorithm yet and therefore its parent in the initial tree $T_k$ also is $v_i$. Since $v_i\neq a$, $v_j\neq v_n$. Since $v_j\notin S_k$, from the definition of $T_k$, $v_i$ is the highest-numbered neighbor of $v_j$. This implies $i>j$, leading to contradiction. Now suppose $i>j$; then $v_j\notin S_k$ must have already been reattached by the algorithm to its lowest-numbered neighbor. This implies $i<j$, yielding contradiction again.

In the second loop, reattachments are undone in the exactly opposite order, except for $v^*_k$; thus, if $v$ is not a leaf in an iteration of the second loop, the only possibility is when $v^*_k$ is its child. However, $u^*_k$, the new parent of $v^*_k$, is in $S_k$, whereas $v\notin S_k$.

Finally, observe that all the above constructions can be performed in polynomial time.
\end{proof}

\section{Lower bound}

Now we exhibit a family of graphs for which the diameter of the graph of spanning trees is $\Omega(|V|^2)$.

\begin{defn}
For $k\geq 1$, $G_k=(V_k,E_k)$ is a graph with $4k+1$ vertices and the specified vertex $a=v_0$, defined as follows:\begin{eqnarray*}
V_k&:=&\{v_0,\ldots,v_{4k}\},\\
E_k&:=&\{(v_0,v_1),(v_0,v_2)\}\cup\\
&&\left( \cup_{i=0}^{k-1}\{(v_{4i+1},v_{4i+2}),(v_{4i+2},v_{4i+3}),(v_{4i+3},v_{4i+4}),(v_{4i+4},v_{4i+1})\}\right)\cup\\
&&\left( \cup_{i=0}^{k-2}\{(v_{4i+4},v_{4i+5}),(v_{4i+3},v_{4i+6})\}\right);
\end{eqnarray*}
$T_k^A$ and $T_k^B$ are its two spanning trees defined by:\begin{eqnarray*}
e_i&:=&(v_i,v_{i+1}),\\
E(T_k^A)&:=&\{e_0,\ldots,e_{4k-1}\},\\
E(T_k^B)&:=&\{(v_0,v_1),(v_0,v_2)\}\cup\\
&&\left( \cup_{i=0}^{k-1}\{(v_{4i+1},v_{4i+4}),(v_{4i+2},v_{4i+3})\}\right)\cup\\
&&\left( \cup_{i=0}^{k-2}\{(v_{4i+4},v_{4i+5}),(v_{4i+3},v_{4i+6})\}\right).
\end{eqnarray*}
\end{defn}

\begin{figure}[h]
\center
\includegraphics[width=.7\textwidth]{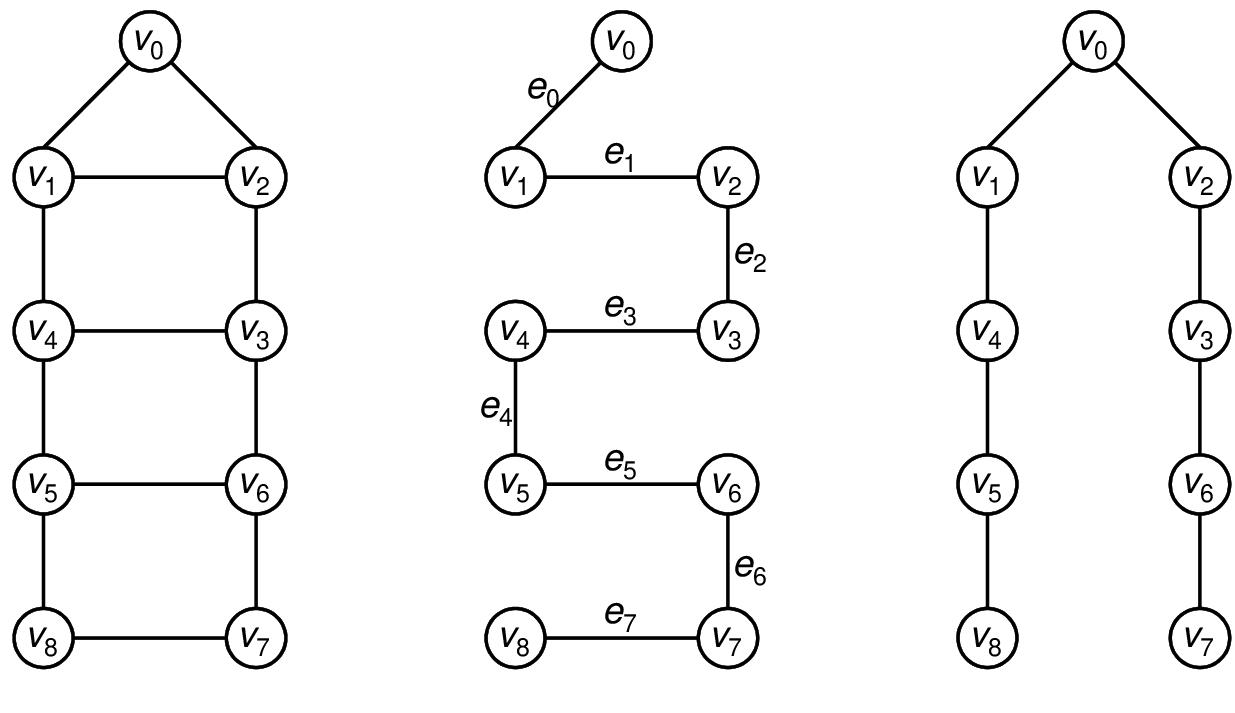}
\caption{$G_2$, $T_2^A$, $T_2^B$.}
\end{figure}

It is easy to observe that $G_k$ is 2-vertex-connected.

\begin{thm}
Let $T_1,\ldots,T_{\ell}$ be a shortest sequence of spanning trees of $G_k$ beginning with $T_1=T_k^A$ and ending with $T_{\ell}=T_k^B$ such that every pair of consecutive trees are adjacent. The sequence length
satisfies $\ell=\Omega(|V_k|^2)$.
\end{thm}
\begin{proof}
Let $t_i$ be the smallest $t$ such that $e_i\in T_t$ and $e_i\notin T_{t+1}$; if there is no such $t$ then $t_i:=\infty$. For $i\neq j$, $t_i\neq t_j$ or  $t_i=t_j=\infty$ since otherwise the intersection of $T_{t_i}$ and $T_{t_i +1}$ contains at most $n-3$ edges. We have $t_1<\infty$ because $e_1\in T_1$ and $e_1\notin T_{\ell}$.

We claim that $\min\{t_0,\ldots,t_i\}=t_i$ for all $i=1,\ldots,4k-1$. Let $t^*:=\min\{t_0,\ldots,t_i\}$. We have $t^*<\infty$ from $t_1<\infty$.
Since $\{e_0,\ldots,e_i\}\subseteq T_{t^*}$ and therefore every endpoint of $e_0,\ldots,e_{i-1}$ either has degree at least 2 or is $v_0$, we have $\{e_0,\ldots,e_{i-1}\}\subseteq T_{t^*+1}$.
This shows $t_0,\ldots,t_{i-1}>t^*$, proving the claim. The claim yields $t_{4k-1}<\cdots<t_1<\infty$.

For $0\leq i<k$, consider $T_{t_{4i+3}}$. Since $t_{4i+2}>t_{4i+3}$, $e_{4i+2}$ is in $T_{t_{4i+3}}$ and $v_{4i+3}$ is not a leaf in $T_{t_{4i+3}}$; $v_{4i+4}$ is a leaf with parent $v_{4i+3}$. Thus, every vertex $v_j$ for $j>4i+4$ must be connected to $v_0$ through $v_{4i+3}$; the subtree rooted at $v_{4i+3}$ contains at least $4k-(4i+3)$ vertices (excluding $v_{4i+3}$ itself). On the other hand, in $T_{t_{4i+2}}$, $v_{4i+3}$ is a leaf (this follows from $t_{4i+1}>t_{4i+2}$ using an argument analogous to the above). Observing that the number of vertices in the subtree rooted at $v_{4i+3}$ decreases by at most one between each consecutive pair of spanning trees in the sequence, $t_{4i+2}-t_{4i+3}\geq 4k-(4i+3)$.

We have $\ell\geq \sum_{i=0}^{k-1} [4k-(4i+3)]=\Omega(k^2)$.
\end{proof}

\bibliography{diameter}

\end{document}